\documentclass[journal,final,finalsubmission, 10pt, twocolumn]{IEEEtran}
\usepackage{amssymb}
\usepackage{amsmath}
\usepackage{graphicx}
\usepackage{array}
\usepackage{algorithm}
\usepackage{algpseudocode}
\usepackage{url}
\usepackage{multirow}
\usepackage{flushend}
\usepackage{engord}
\usepackage{xspace}
\usepackage{textcomp}
\usepackage{verbatim}
\usepackage{hyperref} 
\usepackage[TABBOTCAP]{subfigure}

\usepackage{tikz}
\usepackage[stable]{footmisc}
\usepackage{times}  
\usepackage{epsfig}
\usepackage{tabularx}
\usepackage{color}
\usepackage{thumbpdf}
\usepackage{listings}
\usepackage{booktabs}
\usepackage{colortbl}
\hypersetup{pdfstartview=FitH,pdfpagelayout=SinglePage}

\usepackage[T1]{fontenc}
\usepackage{inconsolata}
\definecolor{bluekeywords}{rgb}{0.13,0.13,1}
\definecolor{greencomments}{rgb}{0,0.5,0}
\definecolor{redstrings}{rgb}{0.9,0,0}

\usepackage{listings}
\usepackage{ifthen}

\usepackage{color}

\usepackage{datetime}

\usepackage{helvet}
\newboolean{draft}
\setboolean{draft}{false}

\ifdraft
\usepackage{calc} 
\usepackage{ifthen} 
\newcounter{hours}\newcounter{minutes} 
\newcommand\printtime{% 
  \setcounter{hours}{\time/60}% 
  \setcounter{minutes}{\time-\value{hours}*60}% 
  \ifthenelse{\value{hours}<10}{0\thehours}{\thehours}:\hspace{-0.33em} 
  \ifthenelse{\value{minutes}<10}{0\theminutes}{\theminutes} 
} 

 \usepackage{draftwatermark}
 \SetWatermarkScale{1.1}
 \SetWatermarkLightness{0.9}
 \SetWatermarkText{\shortstack{DRAFT  \\[0.5cm] {\fontsize{38}{12}\selectfont 
         \today, {\currenttime}}%, \printtime
     }}
  \fi

\def\etal{{\it et al.}}

\begin{document}

\title{Setting the Foundations for PoP-Based\\Internet Evolution Models\IEEEauthorrefmark{1}}

\author{
\begin{tabular}{*{2}{>{\centering}p{.5\textwidth}}}
\large Noa Zilberman & \large Yuval Shavitt \tabularnewline
University of Cambridge & Tel Aviv University \tabularnewline
 Email: {noa.zilberman@cl.cam.ac.uk} & Email: {shavitt@eng.tau.ac.il} 
\end{tabular}
\thanks{\IEEEauthorrefmark{1}This paper was adapted from~\cite{zilberman13thesis}, published in 2013.} 
}

  % make the title area
\maketitle

   %what is the problem?
Developing an evolution model of the Internet has been a long standing research challenge.    
   %Why is it important?
Such a model can improve the design and placement of communication infrastructure, reducing costs and improving users' quality of experience. While communication infrastructure is tightly coupled to geographical locations, Internet modelling and forecasting in the last decade used network elements that are only loosely bounded to any geographical location. 
   %What do we propose?
 In this paper we set the foundations for developing an evolution model of the Internet based on the Point of Presence (PoP) level. As PoPs have a strong geographical grip they can better represent the evolution of the Internet. We annotate the PoP topologies of the Internet with geographical, economic and demographic information to achieve an understanding of the dynamics of the Internet's structure, in order to identify the constitutive laws of Internet evolution.    
   %Why is it good
We identify GDP as the strongest indicator on the country level, and the size of the TV market as the strongest indicator on the US metropolitan level. Finally, we draw attention to the limitations of developing a world-wide evolution model.

\section{Background}
One of the dreams of mankind has always been to be able to predict the future. In scientific terms, this corresponds to the mathematical description of patterns found in real world data in order to devise models that can be used to predict future events. Researchers have pursued a similar goal over the past decade in the area of Internet modeling and forecasting while using Autonomous System (AS) level maps. Efforts have first been focused on obtaining topological maps of the Internet, principally at the Internet Router (IR) and at the AS granularity levels. 
%The AS level representation of the Internet refers to autonomously %administered networks, which are typically corporate networks, ISPs, or %backbone networks. 
In both the IR and AS cases, Internet maps are usually viewed as undirected graphs in which vertices represent routers or ASes and edges represent the physical connections between them.  
Several large-scale measurement projects have started to go beyond these purely topological characterizations of the Internet's properties, and to tackle the characterization and modeling of the relationship between economic factors and Internet evolution. The promised forecast capabilities however have not yet been achieved due to the lack of sufficient data and the difficulty of integrating Internet data with geographical and economic data at a planetary scale.

Internet maps can be presented at several levels, each level of abstraction is suitable for studying
different aspects of the network. The most detailed level is the IP level,
while the most coarse level is the Autonomous System (AS) level.
An interim level of aggregation between the router and the AS level graphs is the PoP level.
A PoP is a group of routers which belong to a single AS and are physically
located at the same building or campus. PoP level maps have been constructed
from various data sources. Andersen \etal~\cite{Andersen02}
used BGP messages for clustering IPs and validated their PoP extraction
based on DNS. Rocketfuel \cite{Spring02journal} generated
PoP maps using tracers and DNS names. The iPlane project \cite{madhyastha08thesis}
generated PoP level maps by first clustering IP interfaces into
routers by resolving aliases, and then clustering routers into PoPs
by probing each router from a large number of vantage points and assuming that 
the reverse path length of routers in the same PoP will be similar.
The DIMES project, takes a structural approach and looks for bi-partite
subgraphs with certain weight constraints in the IP interface graph
of an AS~\cite{FSZ-Comnet12}. The bi-partites serve as cores of the PoPs and are extended with other nearby interfaces.  

The PoP topologies of the Internet used in this paper are annotated with geographical, economic and demographic information to achieve an understanding of the dynamics of the Internet's structure, %at varying time scales,
in order to identify the constitutive laws of Internet evolution. These can be used to develop a realistic topology generator and a reliable forecast framework that can be used to predict the size and growth of the Internet as economies grow, demographics change, and as-yet unattached parts of the world connect. 

The combination of the technological infrastructure with monetary aspects can provide an understanding of the forces driving the data-communications industry today. Using tools and methods from the field of complex science (for example, from statistical physics) it is theoretically possible to develop a prediction model.
The practical uses of an evolution model are numerous: Internet service providers can leverage the model to decide whether to expand their PoP, upgrade its technology or build a new point of presence. City planners can predict its required infrastructure and assign resources for it in advance. Telecommunication firms and semiconductor corporates can better plan their next generation of product and adapt its schedule and features to the market needs. Last, the growth and strength of developing countries can be assessed and predicted, providing country and world level decision makers with essential information in times of economic crisis and market instability.

In this paper, we set the infrastructure for a development of a future evolution model. As we show next, the information required for the development of such a model is yet out of reach. Thus, the following paper surveys the relationships between PoPs and economic and demographic aspects, but only over specific time periods. Using this information, a model can be developed once more information is gathered as years go by.

\section{Related Work}
Learning the dynamics of the Internet and correlating its structure
to drivers in the physical world is important. These drivers may stem from economic incentives, geographic limitations or any other day-to-day life aspect, as was shown in previous works; Many models have been suggested over the years to explain the Internet's evolution, most of them were surveyed and discussed by Pastor-Satorass and Vespignani \cite{PV04}, but there are also later works such as Dhamdhere and Dovrolis~\cite{DD11}, Wang and Loguinov~\cite{WL10} and
Shakkottai \etal~\cite{SFKKC09}.
The models are mostly evolving in the abstract Internet AS graph with no connection to the real world geography, or with some naive connection with the Internet
underlying geography. Some of these works, such as \cite{DS11}, look at the economic aspects of AS level network topology from the ISP's Type of Relationship (ToR) direction.

As time goes by, there is a growing understanding that the evolution of an Internet region should be estimated by tightly correlating the Internet structure with its underlying geography, and the changes in the economic, social, and even political evolution of the region in question. For example, as the economic status
of a developing country improves, it results in a greater demand for Internet connectivity, leading to a growth in the Internet graph related to
this region. There are only a few works in this research direction due to the difficulty of obtaining a good Internet map: Yook \etal \cite{YJB02} compared router, domain and population density in economically developed areas of the world
and indicated that each of the three sets form a fractal with dimension $D_f=1.5\pm0.1$. Combined with  preferential link attachment they proposed an evolution model. Lakhina \etal~\cite{LBCM03} studied the geographical locations of Internet routers and showed that its density varied widely
across the world, but that there is a strong superlinear relationship to population density in economically homogeneous regions. They also showed that the majority of link formation is based on geographic distance, and applied both aspects to the AS graph. Hameed \etal~\cite{HJCS10} used Rocketfuel's \cite{Spring02journal} PoP Topology and combined it with geographical locations based on population density and technology penetration. They validated their results against the published PoPs locations of seven ISPs within the US. M\'{a}tray \etal~\cite{MHLVC12} examined the spatial properties of the Internet topology and routing using Spotter.
They analyzed the direction-dependence of geographic deviations and gave a description of router density in terms of the geographic layout of end-to-end paths. 

The evolution of the Internet and its relationships to geographic and economic factors is also researched in other fields of study, though applying different methods and on a different scale. Roller and Waverman \cite{RW01} studied how telecommunications infrastructure affects economic growth. This work was followed by other works, such as \cite{CFKW11} and \cite{K09} that studied the economic impact of broadband infrastructure on growth. A research by Kolko \cite{K12} studied the relationship between broadband expansion and local economic growth in the US, but surveyed more indicators, such as industry type, population density, employment and income. He found limited economic benefits for local residents stemming from broadband infrastructure. 
A different type of research comes from the field of urban studies, such as Vinciguerra \etal~\cite{VFV10}. They modeled the evolution of infrastructure networks as a preferential attachment process, yet assumed that geographical distance and country borders provide barriers to link formation in infrastructure networks. 

\section{Datasets and Datasets Limitations}\label{sec:evolution_dataset}
Several types of datasets are used in conjunction in this work.
First, we use DIMES's PoPs dataset~\cite{FSZ-Comnet12}. Two PoP level maps are selected, one from 2012, and one from 2010.
These are described below Section \ref{sec:pops_dataset}. PoP level maps from earlier years lack information, either due to the extent of the measurements, 
their accuracy or the lack of geolocation data from that time. Geolocation data changes over the years, as shown in~\cite{SZ_JSAC11}, which may lead to inaccurate PoPs geolocation.

Second, we use the World Bank's World Development Indicators (WDI) \cite{wdi2012} from May 2012. This dataset contains a collection of development indicators, compiled from officially-recognized international sources. It was the most accurate global development data available at the time of measurements, and includes national, regional and global estimates. The dataset is on country level and
it contains indicators such as population and population's growth, GDP, percentage of Internet users and more (total of 1287 parameters per country) on a yearly basis, from 1960 and up till 2012.

Considerable amount of information is gathered from census data. To this end, several census sources are being used. The United States Census Bureau \cite{us-census} provides several types of USA census information. It collects population and housing information every 10 years, conducts an economic census every 5 years as well as smaller surveys and indicators released annually or several times a year. IPUMS \cite{ipums} is a project dedicated to collecting and distributing of United States and international census data. It provides harmonized data for free in a manner that eases that analysis process.

In order to study the effect of transportation infrastructure, we focus on the United States and use the Department of Transportation's Bureau of Transportation Statistics \cite{us-transportation} to retrieve information on highways infrastructure, busiest airports and more. The bureau provides transportation related economic information as well as connectivity and economic factors.
A main source for economic information is the Bureau of Economic Analysis in the US Department of Commerce \cite{us-bea}, which provides information on aspects such as GDP and income. 

The population of cities is obtained from MaxMind's World Cities database \cite{maxmind_worldcities}, which includes information on the population of most of the world's cities as well as their geolocation. We note that some of the
US level databases also include information about the population, but as the size of the population differs from dataset to
dataset due to different definitions of a city or a metropolitan area, we stick to the same population dataset across the entire analysis. 

\subsection{PoPs Datasets}\label{sec:pops_dataset}

Two datasets are used for the validation of the crawling algorithm: one from 2012, and one from 2010, which was selected
as it was carefully studied in~\cite{SZ_JSAC11,FSZ-Comnet12,shavitt2013improving}
and its characteristics are known.
%thus the accuracy of geolocation databases was already evaluated and
%mistakes and anomalies were already identified.
Both datasets use measurements from DIMES~\cite{WebDimes} and iPlane \cite{Madhyastha06}.
We note that the traceroute measurements are performed differently by DIMES and
iPlane, as every DIMES measurement is combined of a train of four traceroute
 measurements, and only the best time of every hop is used for an edge delay calculation. This affects the results beyond a ratio of $1:4$ in the number of measurements. For example, we filter out faulty traceroute hops, such as IP and AS level loops on edge level. Over $170$ million measurements  
are filtered out of the iPlane measurements, while only $61K$ such measurements are filtered from the DIMES data (DIMES filters some of the measurements before adding them to the database).
Due to the differences, edges discovered by DIMES are annotated
with delay information measured only by DIMES, and iPlane data is used to add edges that were not discovered by DIMES. iPlane typically increases the number of discovered edges by \char`\~20\%, but it measures only a small subset of the edges that DIMES discovered.

{\bf 2010 Dataset} 
The dataset is comprised of $478$ million traceroutes conducted in weeks 42 and 43 of 2010, measured by $1308$ DIMES agents and $242$ iPlane vantage nodes.
Five geolocation databases are used for the naive geolocation of the PoPs: MaxMind GeoIP City\cite{maxmind}, IPligence Max \cite{IPligence}, IP2Location DB5\cite{ip2location}, GeoBytes \cite{geobytes} and HostIP.info \cite{hostip}. Two more geolocation services, NetAcuity \cite{netacuity} and Spotter \cite{LMHSCV11}, were tested for the geolocation of PoPs measured by DIMES alone.
The generated PoP level map contains  $4750$ PoPs and $87.3K$ IP addresses in $1697$ different ASes. $4098$ PoPs are discovered using the DIMES data alone. 
We further extend the map by adding universities, research institutes and exchanges points, which were measured by DIMES and iPlane and whose location is known.

{\bf 2012 Dataset} The measurements in this dataset are taken from weeks 19 and 20 of 2012, starting the 6th of May.
$203$ million traceroutes were collected from $988$ DIMES agents and $153$ iPlane vantage points.
Five geolocation databases are used for the naive geolocation of the PoPs: MaxMind GeoIPLite City~\cite{maxmind}, IPligence Max~\cite{IPligence}, HostIP.info~\cite{hostip}, DB-IP~\cite{dbip} and NeuStar's IP Intelligence (formerly Quova)~\cite{neustar}. The generated PoP level map contains  $5215$ PoPs and $98650$ IP addresses in $2636$ different ASes.
This map contains also universities, research institutes and exchanges points, as in the 2010 dataset.

\subsection{Datasets Limitations}\label{sec:evolution_limitations}
As the ultimate goal of the study of PoPs evolution is to come up with a realistic evolution model, the datasets at hand put restrictions and sever limitations on the ability to develop the model over short time periods. First, WDI dataset is only on country level and not on city level, thus it can not be used to the city level modeling intended by this work. International census data is mostly provided on country level and therefore has the same limitation. Census data poses an additional problem, as census is conducted only once every few years (usually five to ten years) and thus does not allow modeling over shorter time periods. As the Internet changes rapidly and technologies emerge and die within a decade, such time frames are not useful.  Large portions of the US datasets are provided on state level, and only partial information is available on metropolitan level.  
The lack of per-city information limits the coverage of PoP's cities in the development of an evolution model over time.

The PoPs dataset limits the development of the model as well. First of all, due to the nature of the PoPs extraction model, it is not possible to tell whether two PoPs of the same AS in the same city are truly separate or are part of the same PoP that was divided due to a missing measurement of an inner link~\cite{FSZ-Comnet12}. For this reason, most of the analysis is based on the number of aggregated PoPs per AS in a given city, with the information on the total number of PoPs and IP addresses within these PoPs observed but largely not used as an indicator. While this may work well in most of the western world, in other regimes the number of competing ISPs is limited or the government controls the communication market (e.g. in Syria~\cite{SZ12_arabian}). In such countries the number of PoPs per city is limited by these external forces and the study of evolution is inaccurate. Last, and most important, there is no ground-truth dataset of PoPs - neither on country nor on city level. This complicates the validation of this work and mostly limits it to information shared by several specific ISPs.

\section{Analysis}

\subsection{The Relation between PoPs and Population}
Points of presence are likely to be closely related to economic factors of their area of residence. For example, areas which are densely populated are likely to have more service providers than small towns. 
We examine here the correlation both at the country level, which was done before (e.g. \cite{YJB02,LBCM03,DS11}), and at the city level, which we are the first to do.

\begin{figure}
\begin{minipage}[b]{1\linewidth}\centering
\includegraphics[width=0.9\textwidth]{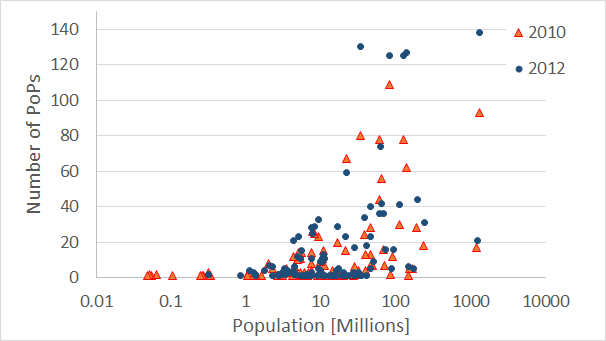}
\caption{Country Level Population vs. Number of PoPs}
\label{fig:pops_population_country}
\end{minipage}
\vspace{-1em}
\end{figure}

\begin{figure}
\begin{minipage}[b]{1\linewidth}\centering
\includegraphics[width=0.9\textwidth]{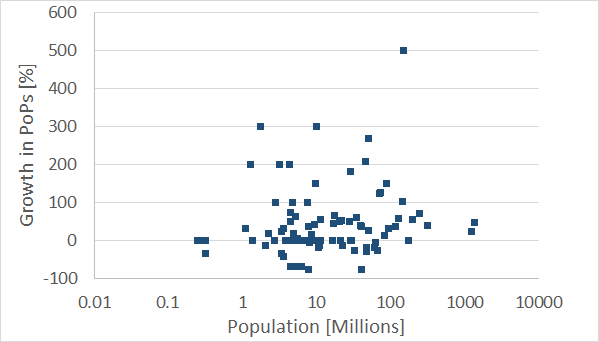}
\caption{Country Level Population vs. Growth in Number of PoPs}
\label{fig:pops_growth_population_country}
\end{minipage}
\vspace{-1em}
\end{figure}

Figure \ref{fig:pops_population_country} shows the number of PoPs discovered on country level compared to the country's population (in millions of people). For clarity, the figure omits the US from the chart, as it is on a different scale.
As can be seen, the size of the population is not a strong predictor for the number of PoPs in a country. The correlation coefficient for population to number of PoPs is 0.22-0.23 both in 2010 and 2012.
To demonstrate this point further, we present in Table \ref{tab:country_population_pops} the number of PoPs per country compared to its population for a set of selected large countries, both for 2010 and 2012.
%PoPs are discovered in over 500 cities for which we have population information. 
The country with most PoPs discovered in 2010 is the US, followed by Germany, China, Canada and Japan.
 In 2012 the list is led by the US, followed by South Korea (Republic of Korea), China, Canada, Russia and Japan. We observe a large
growth in the number of PoPs in South Korea and Japan, whereas in countries such as Germany the number of detected PoPs in 2012 is larger than in 2010, yet in overall it is less than in other countries.
On the other hand, highly populated countries such as India, Indonesia and Bangladesh have very few PoPs. While the number of PoPs does increase between 2010 and 2012, these countries are still lagging behind other large countries. We note that in Pakistan the number of detected PoPs is not only small (5) but also does not change over time.  
On the average, the number of PoPs grew by 38\% between 2010 and 2012 per country, and in 15\% of the countries the number of PoPs doubled itself, as shown in Figure \ref{fig:pops_growth_population_country}. We note that in many of these countries only a handful of PoPs was discovered in 2010. One of the exceptions is South Korea, that had 46 PoPs in 2010 and more than tripled this number in 2012.

\begin{table}
\begin{minipage}[b]{\linewidth}
%\begin{center}
\centering
 \small\addtolength{\tabcolsep}{-1pt}

\begin{tabular}{|l|c|c|c|c|}
 \hline   \bf{} & \multicolumn{2}{c|}{\bf{2010 }} & \multicolumn{2}{c|}{\bf{2012
   }}\\
 \hline
    \bf{Country} & {\bf{Population }} & {\bf{PoPs }} & {\bf{Population }} & {\bf{PoPs}}\\
 \hline
     United States & 306.8M&  850 & 309.3M & 1203 \\
\hline
     South Korea & 49.4M&  46 & 50M & 170 \\
\hline
     China & 1331M & 93 & 1338M & 138 \\
\hline
     Canada & 33.7M &  80  & 34.1M & 130 \\
\hline
     Russia & 141.9M&  62 & 141.9M & 127 \\
\hline
     Germany & 81.9M&  109 & 81.8M & 125 \\
\hline
     Japan & 127.5M &  78 &  127.5M & 125  \\
\hline
     United Kingdom & 61.8M & 78 & 62.2M &  74 \\
\hline
     Australia & 21.9M  & 67 & 22.3M &  59\\
\hline
     Indonesia & 237.4M & 18 & 239.9M &  31\\
\hline
     India & 1207M & 17 & 1224M &  21 \\
\hline
     Bangladesh & 147M & 1 & 148.7M &  6\\
\hline
     Pakistan & 170.5M & 5 & 173.6M &  5\\
\hline

\end{tabular}
\vspace{0.5em}
\caption{Population vs. Number of PoPs on Country Level (Selected Countries) } \label{tab:country_population_pops}
%\end{center}
\end{minipage}
%\vspace{-5em}
\end{table}

\begin{figure}
\begin{minipage}[b]{1\linewidth}\centering
\includegraphics[width=0.9\textwidth]{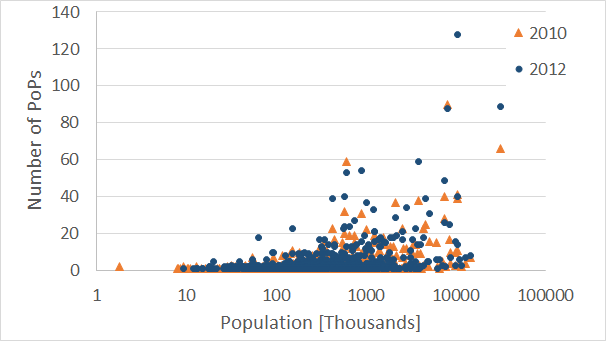}
\caption{City Level Population vs. Number of PoPs}
\label{fig:pops_population_city}
\end{minipage}
\vspace{-1em}
\end{figure}

A different observation is gained by looking at the population versus the number of PoPs on city level. Figure \ref{fig:pops_population_city} shows the number of PoP discovered on city level compared to the city's population and a set of leading cities by PoPs in selected countries is shown in Table \ref{tab:city_population_pops}.
%PoPs are discovered in over 500 cities for which we have population information. 
The city with most PoPs discovered in 2010 is New York, followed by Tokyo, Baltimore, Seoul and London. 
In 2012 Seoul takes the lead with 128 PoPs, followed by Tokyo, New York, Los Angeles and San Jose. In all cases we count PoPs belonging to distinct service providers. In several cases the number of PoPs is decreased between 2010 and 2012, which may be due to lack of measurements, but is also possibly caused by the acquisition or merging of some ISPs. The correlation coefficients for a city's population and the number of PoPs are 0.49 (2010) and 0.51 (2012). 

\begin{table}
\begin{minipage}[b]{\linewidth}
%\begin{center}
\centering
 \small\addtolength{\tabcolsep}{-1pt}

\begin{tabular}{|l|c|c|c|}
 \hline
    \bf{City} & {\bf{Population }} & {\bf{PoPs 2010 }} & {\bf{PoPs 2012}}\\
 \hline
     Seoul & 10.3M&  41 &  128 \\
\hline
     Tokyo & 31.5M\footnote{Refers to Tokyo metropolitan area} &  66 &  89 \\
\hline
     New York & 8.1M&  90 &  88 \\
\hline
     Los Angeles & 3.9M & 38 &  59 \\
\hline
     London & 7.4M&  40 &  49 \\
\hline
     Moscow & 10.4M &  39  &  40 \\
\hline
     Shenzhen & 10M &  22 &   37  \\
\hline
     Paris & 2.1M &  37 &   29  \\
\hline
\end{tabular}
\vspace{0.5em}
\caption{Population vs. Number of PoPs on City Level (Leading Cities in Selected Countries) } \label{tab:city_population_pops}
%\end{center}
\end{minipage}
%\vspace{-5em}
\end{table}

We study the inflation in the number of PoPs in Seoul and find that there are two reasons for that.
First, recall that all AS-level PoPs are aggregated on city level, thus we do not count the same autonomous system
more than once. When examining the active ASes located in Seoul we find that most of them belong to universities:
While in 2010 only three detected ASes belonged to universities, in 2012 there are 56 such ASes. ASes that belong to 
other educational and technological institutes are added on top of that, leading to a total growth from 15 to 83 PoPs.
While these results raise suspicion regarding the accuracy of the PoPs' geolocation and possible mistakes in geolocation databases, this was found not to be the case, except for a few cases that apply to Seoul's suburbs. It was manually corroborated that the results are true and that the increase in the number of PoPs was as a result of a government policy\footnote{We'd like to thank Dr. Jong Hun Han for his assistance on this subject}.

In most of the capitals and large cities of the developed world, tens of PoPs are detected, but in some 
of the most populated cities of the world, such as Bombay, Manilla and Delhi only a handful of PoPs are detected.
While the number of PoPs discovered in these cities grows between 2010 and 2012, it does not significantly change:
in Bombay the number of PoPs grew from 4 to 7, in Delhi from 2 to 6 and in Manilla from 9 to 14. In comparison, the number of PoPs in Seoul grew from 41 to 128 and in Los Angeles from 38 to 59. While one may attribute this to the
number of Internet users in a country, correlating the number of Internet users or the percentage of Internet
users in a country to the number of PoPs is not a good indicator either (see Section \ref{sec:evolution_users}). In addition, while the number of PoPs depends on
the number of measurements in a target country, in practice this effect is small, as DIMES and iPlane try to reach all possible IP prefixes.
PoPs are also detected in small cities, such as Larnaca, Cyprus (less than 50K inhabitants).

Comparing the number of PoPs to the population is somewhat misleading, as countries are considerably different from each other, and one can not compare, for example, the United States to Bangladesh based on population alone.
For this reason, we break down city level analysis and conduct it per country. We include in the analysis only countries where PoPs were detected in at least 5 cities, and check whether the number of PoPs in a city corresponds 
to the city's size. The dataset, based on 2012 PoP level map, includes 24 such countries and 508 cities. For each country we rank the cities by population
and check if the ranking by PoPs' number is identical. We find that for 15 countries the rankings of population and PoPs match,
assuming that we allow up to two PoPs difference, since it is negligible. A different view on this aspect is gained by binning.
The number of PoPs per city is divided to three bins: 5 PoPs or less, 6-10 PoPs and more than 10 PoPs. Cities are also divided to
small cities (100K residents or less), medium cities (100K-1M residents) and large cities (1M residents or more). We find that using 
this binning, 21 of the 24 countries have a full match between the ranking of PoPs and the population's ranking. This means that on a country level, the size of a city is an excellent indicator to the number of PoPs in it, but the ratio between the number of PoPs and population varies between countries. The three countries that do not match this observation are the United States, Italy and Germany.
In Italy, Bologna has 6 PoPs, while in larger cities like Turin and Naples only 3-4 PoPs are detected. In Germany, significantly more PoPs are discovered in Frankfurt (24) compared to Berlin (4) and Hamburg (6) despite the latter being more than twice its size. In the US many such cases exist, possibly because in many cases the PoP is located in a small town close to a large city. While the anomalies can be explained by other factors, the population is shown not to be the only indicator to determine the number of PoPs per city.

\subsection{The Relation between PoPs and GDP}

\begin{figure}
\begin{minipage}[b]{1\linewidth}\centering
\includegraphics[width=0.9\textwidth]{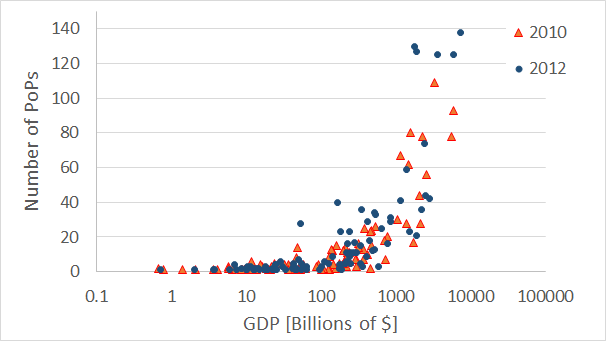}
\caption{Country Level GDP vs. Number of PoPs}
\label{fig:pops_gdp}
\end{minipage}
\vspace{-1em}
\end{figure}

The GDP of a country is a good indicator to its number of PoPs. There is a clear relation between
the GDP and  the number of PoPs, as shown on Figure \ref{fig:pops_gdp}. The figure shows on 
country level the number of PoPs per country compared to its GDP for 2010 and 2012 datasets; for the 2012
dataset we used the GDP reported at the end of 2011 (as published in WDI's May-2012 dataset).
As the figure shows, high GDP leads to a high number of PoPs on the country level. The correlation coefficient between the GDP and number of PoPs is very high: 0.92 in 2010 and 0.90 in 2012. For countries with a GDP of 100's of billions of dollars, this is clearly the trend, but it is not always the case. For example, Sweden and Saudi Arabia have almost the same GDP (538 and 577 billions of dollars, respectively) yet in Sweden we detect in 2012 thirty three PoPs, while only three PoPs are detected in Saudi Arabia.
For this reason, a simple equation that shows the relation between the GDP and number of PoPs can not be found without a meaningful square root error. Both 2010 and 2012 datasets exhibit a similar pattern and are overlapping in many points.

\begin{figure}
\begin{minipage}[t]{1\linewidth}\centering
\includegraphics[width=0.9\textwidth]{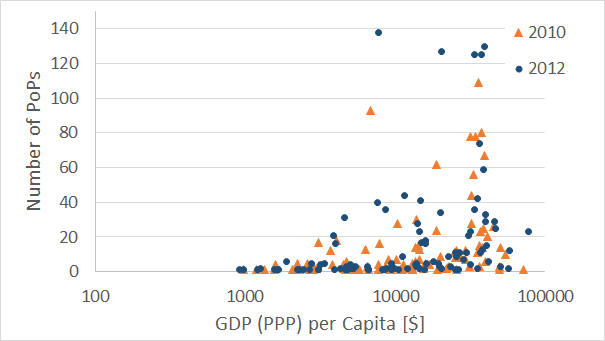}
\caption{Country Level GDP (PPP) per Capita vs. Number of PoPs}
\label{fig:pops_ppp}
\end{minipage}
\vspace{-1em}
\end{figure} 

\begin{figure}
\begin{minipage}[t]{1\linewidth}\centering
\includegraphics[width=0.9\textwidth]{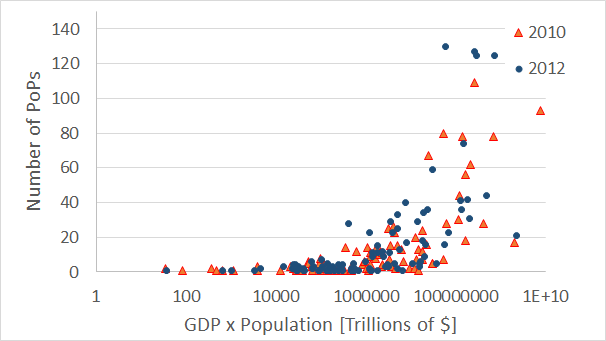}
\caption{Country Level GDP x Country Level Population vs. Number of PoPs}
\label{fig:pops_gdp_population}
\end{minipage}
\vspace{-1em}
\end{figure}

One may expect that other types of predictors relating to GDP will provide a better indication for the number of PoPs.
One such parameter is the GDP (PPP) per capita, meaning the gross domestic product at purchasing power parity per person, which is often considered an indicator of a country's standard of living. However, this turns out not to be a good indicator, as shown in Figure~\ref{fig:pops_ppp}: some countries have very high GDP per capita but very few PoPs (e.g. Qatar, Kuwait) whereas countries such as China and Russia have considerably lower GDP per capita, but many more PoPs. The correlation coefficient in this case in only 0.25, both in 2010 and 2012.

Another possible predictor for the number of PoPs, complementary to the previous one, is the multiplication of GDP in the population, however this turns out to yield results that are slightly less aligned with the best fitted linear line than the dependence on GDP alone, as shown in Figure~\ref{fig:pops_gdp_population}, but more aligned than PPP. The correlation coefficient in this case is 0.55 in 2010 and 0.5 in 2012. This indicator may explain why countries with high GDP and small population have the same number of PoPs as countries with a large population but a medium GDP: the GDP is not the only factor, so very large countries with a medium 
GDP will still need a significant PoPs infrastructure, to provide Internet services its residents.

The growth in GDP is not an indicator to the number of PoPs, and countries with high GDP growth do not have more PoPs than countries with low or negative GDP growth. The correlation coefficient here is neutral: ranging from zero to $-0.05$. Similarly, the growth in the number of PoPs between 2010 and
2012 is not correlated with the growth in GDP. An example to this is Japan, that had at the end of 2011 a GDP growth of $-0.7\%$
whereas its number of PoPs grew by 60\%.

\subsection{The Relation between PoPs and Internet Users}\label{sec:evolution_users}
At a first glance, the number of Internet users per country may seem a good indicator for the number of PoPs: one may expect that the need for PoPs will rise as more Internet users require Internet connectivity.
This assumption, however, if not founded. When studying the relationship between the number of PoPs and the number of Internet users
per country\footnote{This study is limited to the 2010 dataset only, as the WDI dataset included only the number of Internet users up till 2010, and for consistency we chose not to take this data from later datasets.}, there is some weak relation in countries with many Internet users. Meaning, most countries with 20 PoPs or more have tens of millions of Internet users. Yet this is not always the case: countries such as Austria, Sweden and Switzerland have six to 
eight million Internet users, but over 25 PoPs in each. Considering the other way around, i.e., whether a large number of Internet users
calls for a large number of PoPs, there are some exceptions as well: Nigeria, Turkey and Pakistan all have twenty eight million Internet
users or more, but five PoPs or less. The correlation coefficient between the number of Internet users and the number of PoPs is 0.53 in this case. 

One explanation may stem from the percentage of Internet users in the population (information taken from the WDI dataset), but our
analysis shows no connection between the percentage of Internet users and the number of PoPs. It is not only weaker than the total
number of Internet users versus PoPs, it seems to be merely related, with a correlation coefficient of only $0.18$. The same applies also for the average bandwidth per user, where the correlation coefficient is $0.19$ (based on~\cite{akamai12} and covering 49 countries).

A possible explanation to why the number of PoPs does not depend on the number of Internet users, is that service providers not necessarily
have to increase the number of PoPs in order to handle increasing demand for Internet access. For example, they can expand existing PoPs,
adding more networking equipment and thus exposing more ports towards the end users. The providers can also replace the technology used in their PoP, e.g., using 10GE interfaces instead of 1GE. Last, it is possible that in dense areas, such as crowded cities, we fail to detect 
multiple PoPs per a single ISP, due to the nature of our algorithm.

\subsection{A Study of the United States}
The United States is a special case amongst all countries. First of all, the number of PoPs detected in it is extremely high (1203 in 2012). Second, it is a vast country, with a high GDP, considerable population and it is technologically advanced. Last, the large amount of economic and demographic information which is available on city and metropolitan level, enables us to perform more accurate and advanced studies.

\subsubsection{The Relation between PoPs and Population in the US}\label{sec:evolution_us_population}
\begin{figure}
\begin{minipage}[b]{1\linewidth}\centering
\includegraphics[width=0.9\textwidth]{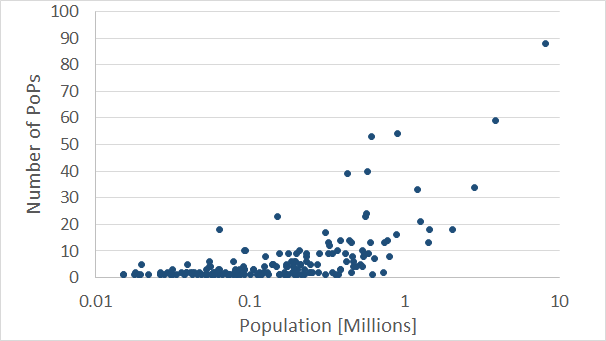}
\caption{United States City Level Population vs. Number of PoPs}
\label{fig:pops_city_population_us}
\end{minipage}
\vspace{-1em}
\end{figure}

As noted above, the US is one of three countries where no direct relation is observed between city level population and the 
number of PoPs in that city. Figure \ref{fig:pops_city_population_us} demonstrates this, with the X-axis being the number of residents in a city (in Millions)
and the Y-axis being the number of PoPs (aggregated by AS) in that city.  The correlation coefficient, which is 0.79 and 0.78 in 2010 and 2012, correspondingly, does not tell the whole story: While for many cities, like New York and Los Angeles,
the rule that more residents mean more PoPs applies, there are many exceptions. Amongst the medium-size cities (less than a million people)
one can find cities like Boston or Baltimore with 500K to 600K residents but over 40 PoPs. Many PoPs are sometimes found in small cities as well: 23 PoPs in Springfield, MO (150K residents) or 10 PoPs in Albany, NY (94K residents). %This is %partly the effect of observing the PoPs in over 160 cities. 
Consequently, additional indicators need to be found for the number of PoPs in a city.

The study of additional indicators uses metropolitan level statistics, rather than city level, as  this level of aggregation has the most information from official US government sources, such as the Bureau of Economic Analysis (BEA), Bureau of Transportation Statistics (BTS) and above all the US Census Bureau. On the PoP level the usage of metropolitans rarely affects the results due to range of convergence applied when assigning PoPs to cities. In a handful of cases where a metropolitan area includes more than a single city, such as Dallas-Port Worth, we aggregate the
PoPs' city level information to the metropolitan level.

\subsubsection{The Relation between PoPs and GDP in the US}
\begin{figure}
\begin{minipage}[b]{1\linewidth}\centering
\includegraphics[width=0.9\textwidth]{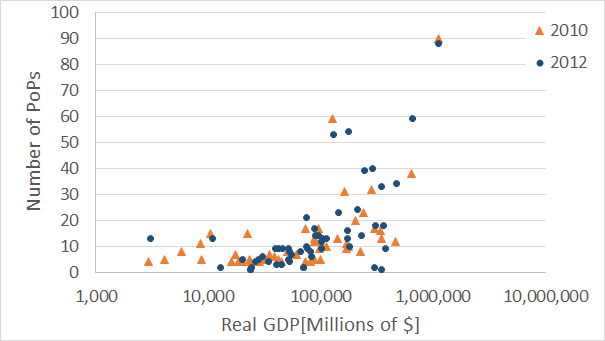}
\caption{United States Metropolitan Level GDP vs. Number of PoPs}
\label{fig:pops_city_gdp_us}
\end{minipage}
\vspace{-1em}
\end{figure}

One economic aspect that was studied on country level and can now be observed on metropolitan level is GDP.
We study the real GDP (in millions of chained 2005 dollars) as provided by the BEA \cite{bea-gdp}, as a total of all industries. The analysis covers the 50 largest metropolitans (by population) in which we detected PoPs in 2010 and 2012.
As opposed to what one may expect, the correlation between GDP and number of PoPs in weaker on metropolitan level: only 0.78 in 2010 and 0.76 in 2012. While the correlation is still evident, it is not as strong as on the country level. This is demonstrated in Figure \ref{fig:pops_city_gdp_us}. The GDP is shown in millions of dollars on the x-axis, whereas the number of PoPs is shown on the y-axis. While in most metropolitan areas the change in GDP between the years is small, while the number of PoPs rises, there are a few metropolitans where the number of PoPs decreases. Although this may be attributed to lack of measurements, this is also the result of acquisition or merging of some ISPs, which cause a convergence of PoPs in a given area, as we count the PoPs of every AS only once per city.

\begin{figure}
\begin{minipage}[b]{1\linewidth}\centering
\includegraphics[width=0.9\textwidth]{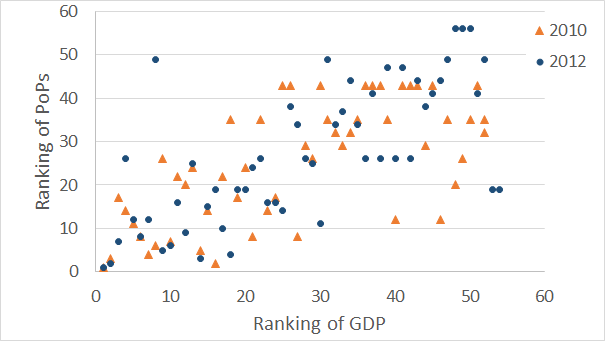}
\caption{United States Metropolitan Level Ranking of GDP vs. Ranking of PoPs}
\label{fig:pops_city_gdp_ranking_us}
\end{minipage}
\vspace{-2em}
\end{figure}

Another way to consider the relation between GDP and PoPs is using ranking: We rank the metropolitans by the number of PoPs in them, with the highest rank going to the metropolitan with most PoPs and the lowest rank to the one with least PoPs. If two metropolitans have the same number of PoPs, their ranking is similar. Identical ranking is applied to each metropolitan's GDP. This method is selected as much of the US metropolitan area statistics is
published using ranking. Figure \ref{fig:pops_city_gdp_ranking_us} demonstrates the relation between the GDP ranking and the ranking of PoPs: generally speaking, the higher the GDP of a metropolitan, the higher its PoPs' ranking. However, this is not
a clear linear relation and there are some exceptions, e.g. San Francisco, ranked 8th by GDP but only 48th by PoPs (in 2012).
The correlation coefficient in this case is another evidence: In 2010 it is 0.65 and in 2012 it is 0.71. Both coefficients are weaker than the correlation coefficients between the Real GDP and the number of PoPs.  

 Another economic factor that is considered is personal income: we study the per capita personal income in the same metropolitan areas, as published by the BEA \cite{bea-income}.  The correlation between income and the number of PoPs is weaker than GDP, yet stronger than the country level PPP; It reaches 0.5 in 2012 and 0.63 in 2010. The large gap between the two datasets is another reason not to consider this parameter as a good indicator. 
 
\subsubsection{The Relation between PoPs and Transportation in the US}
It is a common assumption that networking infrastructure is tightly related with transportation
infrastructure \cite{yoshida09}, such as railways and highways, and that main transportation hubs
also serve as communication hubs. We examine this assumption when considering PoPs and various transportation related statistics in the United States.

The first aspect under study is the US' top freight gateways, in sea, air and land \cite{bts-gateways}. As these gateways require significant infrastructure in order to transport the cargo, it is interesting to check whether the same locations also serve as  networks' landing points and as centers of PoPs. The dataset is compared to 2012 PoPs dataset. It is found that only twenty of the metropolitans under study are included within the Top 50 freight gateways (the size of the dataset). In this count we include also gateways that are in proximity (up to 150km) of the metropolitan.  Four of the metropolitans have more than a single type of a gateway in the list: Houston, Los Angeles, Miami and New Orleans. Ten of these gateways are through water, twelve through air and only three through land. The calculated correlation coefficient between the total value of shipments through a gateway and the number of PoPs is 0.58, and only 0.34 when the ranking of a gateway and the ranking of PoPs is considered. This calculation uses only the small set of twenty metropolitans, and is thus very sensitive and prone to fluctuations.  

Following freight gateways, we focus on passengers transportation through airports. A database of top 50 US airports is used for this end \cite{bts-airports} and is compared with the 2012 PoPs dataset. Out of the top 50 airports, 37 are included in the top 50 metropolitans with PoPs, and a total of 41 in the list of all cities with PoPs. While this indicates that metropolitans with considerable air traffic are likely to have a lot of active ISPs, the relation between the number of PoPs and the amount of passenger is weak: a correlation coefficient of 0.3. The relation between the ranking of a metropolitan by PoPs and by airport's passenger is even weaker, 0.11.

Another type of transportation infrastructure is denoted by railways. While not a direct indicator of railway tracks infrastructure, we examine the top 50 Amtrak stations by number of passengers and compare it to the 2012 PoPs dataset.
Just 15 metropolitans are shared between the list of top metropolitans with PoPs and top Amtrak stations. Six more metropolitans appear
in the full list of metropolitans with PoPs. However, one needs to note that the Amtrak ranking list includes some cities
more than once, thus there are only 43 distinct metropolitans in it. Fifteen overlapping metropolitans can be considered insufficient to calculate the correlation coefficient, but for illustration purposes, we find it to be 0.44 for the relation between number of passenger and number of PoPs and 0.37 between a station's ranking and the PoPs' ranking.

\begin{figure}
\begin{minipage}[b]{1\linewidth}\centering
\includegraphics[width=0.9\textwidth]{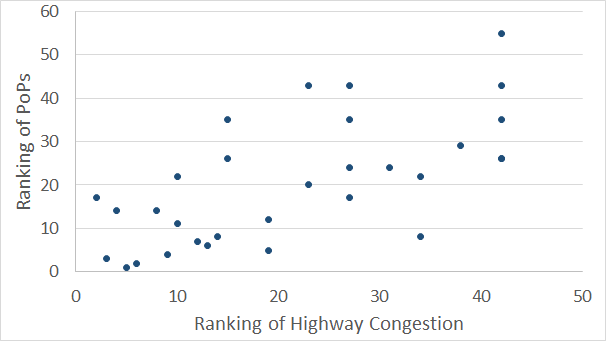}
\caption{United States Metropolitan Level Ranking of Highway Congestion vs. Ranking of PoPs}
\label{fig:pops_congestion_ranking_city_us}
\end{minipage}
\vspace{-2em}
\end{figure}

The last case relating to transportation under study is highway congestion in the 50 largest urban areas \cite{bts-congestion}. This set matches 35 metropolitans in the 2010 dataset\footnote{We use the latest BTS dataset available at the time.}. Surprisingly, we find here better correlation to the number and ranking of PoPs compared to the previous cases:
the correlation coefficient between the total hours of delay and the number of PoPs is 0.61 and the correlation coefficient between the ranking of a metropolitan by a highway congestion delay per commuter and its ranking by PoPs is 0.66. While this is not a strong correlation as with GDP, it is better than for other types of transportation indicators. The correlation between the  highway congestion delay hours per commuter and the number of PoPs is weaker, being 0.42. The relationship between a metropolitan ranking by highway congestion per commuter versus ranking by PoPs is shown in Figure \ref{fig:pops_congestion_ranking_city_us}. Note that both ranking lists have several metropolitans with the same ranking, due to identical number of PoPs or identical hours of delay per commuter, which affects the correlation and is reflected in the graph.

\subsubsection{The Relation between PoPs and Demographic Factors in the US}
The United States census information is used to study the relation between two demographic aspects and PoPs: age and race.
These two aspects are selected due to their availability, compared to other important aspects that are either not
covered on metropolitan level or that were already covered before in this work, such as income.

The 2010 US census information is used in conjunction with the 2010 PoPs dataset to study the relation between different age groups in a metropolitan \cite{us-age} and the number of PoPs in the area. %Four age groups are considered: Under 18 years,	18 to 44 years,	45 to 64 years,	over 65 years, and the size of their population in a given metropolitan is compared with the number of PoPs. 
Table \ref{tab:table_pops_age_city} shows the correlation coefficient between each age group and the number of PoPs. As can be seen, the correlation coefficient is very similar for all age groups, which may indicate a relation to all age groups. However, as there is a correlation of over 0.99 between the overall size of the population and the size of a specific 
age group, the results actually reflect the relation between PoPs and population as studied in the beginning of this section.
%We note that correlation coefficient to the overall population is not identical to the one stated above, which is a result of the different definition of the city and metropolitan area between the datasets, as explained in Section \ref{sec:evolution_dataset}.

\begin{table}
\begin{minipage}[b]{\linewidth}
%\begin{center}
\centering
 \small\addtolength{\tabcolsep}{-3pt}

\begin{tabular}{|l|c|c|c|c|c|}
 \hline
    \bf{Age Group} & {\bf{All }} & {\bf{Under 18 }} & {\bf{18-44 }} & {\bf{45-64 }} & {\bf{Over 65}}\\
 \hline
     \bf{Correlation}  & 0.76&  0.74 & 0.76 & 0.77 & 0.77 \\
     \bf{Coefficient}  & & & & & \\
\hline

\end{tabular}
\vspace{0.5em}
\caption{Correlation Coefficient Between Size of Different Age Groups and The Number of PoPs } 
\label{tab:table_pops_age_city}
%\end{center}
\end{minipage}
%\vspace{-5em}
\end{table}

To study the relation between race and the number of PoPs, we use the information gathered on the same 2010 US census with the 2010 PoPs dataset. The census dataset \cite{us-race} states for each metropolitan statistical area the number of people by race. Race may be White alone, Black or African American alone, American Indian or Alaska Native alone, or Asian alone. We do not refer to Native Hawaiian and Other Pacific Islander alone as the size of their population is negligible in most metropolitans (a few hundreds of people). In addition, a person may define himself as from two or more races. An exception is people from Latino or Hispanic origin, who may be of any race, and are therefore counted separately (i.e. both under their race and origin). The results portray a complicated story: The correlation coefficient between the size of the population of white, Asian or people of two or more races is almost identical to the correlation coefficient for the entire population. For white people, who are about 78\% of the population this is understandable, as the size of their population has 0.99 correlation coefficient  to the entire population. Asian and people of two or more race, each pose about 2.7\% of the population and have 0.91 and 0.96 correlation coefficient to the overall population, which is weaker than for white people but still very high. The size of the American Indian population has a correlation coefficient of only 0.51 to the number of PoPs, yet they are only 1\% (on the average) of the overall population with 0.74 correlation coefficient to it, so their case might not be well represented. The African American and Hispanic population are a different case: their share of the population is rather large (each over 10\%) and while their correlation coefficient to the overall population is lower than White people or people of two races or more, it is almost the same as that of the Asian population. Yet, the correlation coefficient of these two group is lower by 12\%--18\% compared to other major races. This may still be attributed to the lower correlation between the size of this population and the overall population, but it may also be driven by other social and demographic factors. 
 	
\begin{table*}
\begin{minipage}[b]{\linewidth}
%\begin{center}
\centering
 \small\addtolength{\tabcolsep}{-3pt}

\begin{tabular}{|l|c|c|c|c|c|c|c|}
 \hline
    \bf{Race} & {\bf{All }} & {\bf{White }} & {\bf{African American }} & {\bf{American Indian }} & {\bf{Asian}} & {\bf{Two Races}} & {\bf{Hispanic}}\\
 \hline
     \bf{Correlation}  & 0.76&  0.76 & 0.65 & 0.51 & 0.76 & 0.74 & 0.62 \\
     \bf{Coeff to PoPs}  & & & & & & &\\
 \hline
     \bf{Correlation}  & &  0.99  & 0.87 & 0.74 & 0.91 & 0.96 & 0.89 \\
     \bf{Coeff to All}  & & & & & & &\\
 \hline
     \bf{Average \%}  & &  78.1\% & 10.7\% & 1.0\% & 2.7\% & 2.7\% & 12.4\% \\
     \bf{of All}  & & & & & & & \\
\hline

\end{tabular}
\vspace{0.5em}
\caption{Correlation Coefficient Between Size of Different Race Groups and The Number of PoPs } 
\label{tab:table_pops_race_city}
%\end{center}
\end{minipage}
%\vspace{-5em}
\end{table*}

\subsubsection{The Relation between PoPs and TV Market in the US}
Nielsen Media Research releases every year a rating of Designated Market Areas (DMA) across the US. The size of a market
is measured by the size of the television audience in it, where the audience do not need to live within a city to be considered part of its DMA, rather live where its stations are watched the most. For example, the Philadelphia DMA includes southern New Jersey and most of Delaware. We take Nielsen's 2011-2012 ranking and compare it to the 2012 PoPs dataset.

 \begin{figure}
 \begin{minipage}[b]{1\linewidth}\centering
 \includegraphics[width=0.9\textwidth]{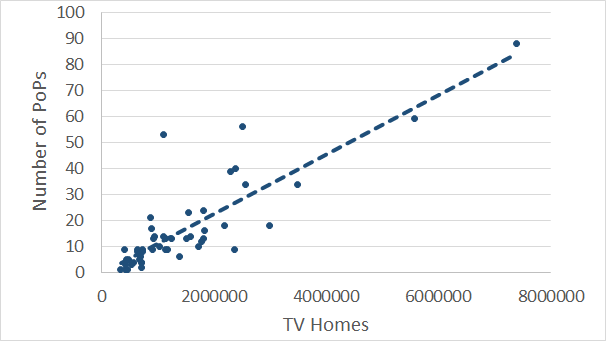}
 \caption{United States Number of TV Homes vs. Number of PoPs}
 \label{fig:pops_tv_city_us}
 \end{minipage}
 \vspace{-2em}
 \end{figure}

As the Nielsen dataset includes not only the ranking of the markets, but also their size by the number of TV homes, we first check the correlation between the size of a TV market and the population of the given DMA. A very high correlation may cause the results to mirror the correlation between PoPs and population and thus make the size of the TV market a redundant indicator. The resulting correlation coefficient is 0.88 (compared with the population dataset used in Section \ref{sec:evolution_us_population}), which is high but does not mean that the relationship is identical. The correlation coefficient between the number of TV homes and the number of PoPs in a city is found to be 0.85, whereas the correlation coefficient between the ranking of a TV market and the ranking of its PoPs is 0.82. Both coefficients are higher than 0.78, which is the 
correlation coefficient for the relation between population and PoPs. Figure \ref{fig:pops_tv_city_us} shows the relationship between the number of TV homes and the number of PoPs, with the dashed line showing the linear relation between the two. The line's coefficient of determination is 0.71. One can expect that the high correlation between the size of the TV market and the number of PoPs will be a result of IPTV penetration, which requires such network infrastructure, however in 2012 IPTV had only 9.66 million subscribers in the United States \cite{tv-penetration}. On the other hand, the penetration of IPTV to broadband users in the country was almost 40\% in 2012 \cite{broadband-penetration}. It thus seems that the right course is to study further the relation between broadband penetration and the number of PoPs. Unfortunately, we did not manage to locate this information on the city or metropolitan level.

\subsubsection{The Relation between PoPs and Sports Teams in the US}
To continue the line presented in the previous subsection, one possible driver of TV markets is sports events,
such as NFL games, that often lead the TV shows ratings lists. It was thus suggested that cities hosting such events
require considerable infrastructure in order to support the media, and therefore it may relate to the number of PoPs.

The major sports leagues in the United States and Canada are the MLB (baseball), NBA (basketball), NFL (football) and NHL (hockey). These four leagues are often called "The Big Four". Adding also the MLS (soccer) and CFL (Canadian football) is referred to as "The Big Six". The Big Six sports teams are located in forty metropolitan area in the US, and 9 more in Canada. We focus on the US sports teams and use the current allotment of teams to a metropolitan area. Information is collected from the official websites of the leagues. We note that since there are no CFL teams in the United States, only MLS teams make the difference between the Big Four and Big Six teams' count.

Out of the forty metropolitan areas where teams are located, we detect PoPs in thirty seven places. The average number of PoPs in each of these metropolitans is 21.1, with the median being 14, and the minimum number of PoPs being 4 in Oklahoma (which has only one sports team, in the NBA). The correlation coefficient between the number of PoPs and the number of Big Four sports team is 0.84 and the correlation to the number of Big Six teams is 0.86. This is a high level of correlation, especially as the correlation between the number of sports teams and the size of the population is no more than 0.7.
%less for big 6
To support this result, the average number of PoPs detected across the entire dataset is 6.5 PoPs per metropolitan area, with the median being 2.5 PoPs and the minimum a single PoP. Considering the group of metropolitans with no sports teams, 
the average number of PoPs is 2.9 and the median is two. The large gap in the number of PoPs between the group of metropolitans with and without sport teams points that this may be a valid indicator. 

While the results above suggest studying the relation between PoPs and National Collegiate Athletic Association (NCAA) teams, 
such a case will be complex: there are hundreds of NCAA teams and the games are on multiple broadcast networks as well as on local TV networks. This indicates that the relation will involve not only the PoPs and number of local teams, but also other factors such as the team's division and ranking, and possibly its market size.

\section{Discussion}
The analysis of the PoP level map versus the various economic and demographic aspects teaches us a few lessons.
The most important lesson is that global analysis is too coarse to lead to a model or a set of indicators
that will apply to all countries. The difference between countries is too large to expect that if a rule
applies to the a country in North America or Europe it will also apply to Africa and Asia. The differences
stem not only from the country's level of development or economic status, but also from government policy - 
for better (South Korea) or worse (Middle Eastern countries \cite{SZ12_arabian}). An evolution model that
will try and predict PoPs evolution over time will therefore need to apply different metrics to different
types of countries. These results corroborate previous works, such as Lakhina \etal~\cite{LBCM03}, which showed
that the number of router interfaces can't be correlated on a worldwide level, but that there is a correlation 
within economically homogeneous regions. 

The analysis of the PoPs on US metropolitan level is in many ways more fruitful than on the country level.
While this is largely due to the availability of information on metropolitan level, the vast size of the country and the large amount of PoPs detected in it, this is also due to derivative cultural aspects. The combination of 
the advanced technological status of the US with leisure culture make the effect of aspects like size of TV market
and sports teams larger than in other places. If one would like to compare these aspects to other comparable areas,
e.g., the European Union, he may find that it is hard. For example, in the TV market each country in the EU may
have its own policy and use a local language, which is different than the US. Sports teams are also managed differently
(e.g., the Football's Champions league and the basketball's Euroleague), as the sports teams included and excluded vary
each year based on local achievements and thus do not form a constant set of teams that requires long-term investment in communication infrastructure 
(except for a handful of leading clubs).

The indicator that turns out to be the most influential on the number of PoPs is the GDP: strongly
on the country level and considerably on US metropolitan level. While analyzing the reason for that 
compared to other economic factors is outside our field of expertise, there is no doubt in the implications
on the number of detected PoPs. On the US city level, the size of the population is a strong indicator to 
the number of PoPs, and it also has a correlation to other aspects with high level of correlation to the number
of PoPs, such as age groups and TV market. 

An unexpected result, from our point of view, was the low level of correlation between different aspects of
transportation and the number of PoPs on the US metropolitan level. As network infrastructure is considered 
related to transportation infrastructure (e.g. \cite{RPK01} and \cite{VFV10}), one would expect higher correlation
between the two. However, since the parameters that we study were limited due to datasets availability, they may not necessarily reflect the entire scope of
transportation infrastructure in an area, which may be the cause for the results.

The lack of information for an evolution study of the PoPs level graph has two contributors. First, the 
economic, geographic and demographic information that is not always accessible and or not available on the required 
points in time, as discussed in Section \ref{sec:evolution_limitations}. The second part is the short history
of the Internet and the radical changes the network has gone through in the recent decade. As information 
from other fields of study has a long history and is commonly sampled on a decade and half decade basis, there
are not enough overlapping sampling points to devise a reliable model based on measured data. 
The datasets used in the analysis are only a year and a half apart and thus are not far enough apart to indicate
growth or change trends over time. While we do detect more PoPs in 2012 than in 2010, it will be incorrect to 
deduce anything based on these differences. The use of the two datasets does support the results regarding the
correlation between different indicators and the number of PoPs, as the correlation coefficients are very
similar across the years.

\section{Conclusion}

In this paper, we set the foundations for PoP-level Internet evolution models.
We examined different aspects of geographic, economic and demographic factors and
explored their correlation to the number of PoPs in different countries and cities around the world.
The results show that GDP is a good indicator of the number of PoPs, on the country level,
and that on the US metropolitan level the population may be a good indicator, as well as several
other related aspects. 

A future evolution model will need to take into account multiple PoP-level maps
spread across a longer span of time, in order to achieve a better understanding of the dynamics
of evolution over time. The model should look on the city level, where possible, and should distinguish
between countries based on their different characteristics (e.g. economic region).
Further work should also involve researchers from other disciplines, such as geographic and economic
studies, in order to better analyze the data and have a better understanding of the results.
A different direction is to perform a focused study of other countries than the United States, and to check
similarities and differences.

Another aspect that we hope to study in the future is the evolution of the PoP level map's technological infrastructure. This means that one should look not only at the number of PoPs but also at the technology that is used in them, e.g., 10GbE, 40GbE or 100GbE, and the number of exposed interfaces in each PoP. This kind
of study will require collaboration with service providers, as the type of infrastructure used is rarely revealed. 
This type of study may better reflect some changes in the evolution of the network, due to the dominance of 
some tier-1 ISPs, who may have greater influence on the network than the introduction of new ones to a city's PoP level map.

\bibliographystyle{abbrv} 
\begin{small}
\bibliography{cites}

\begin{thebibliography}{10}

\bibitem{neustar}
{Neustar}.
\newblock http://www.neustar.biz/, 2012.

\bibitem{us-census}
{United States Census Bureau}.
\newblock http://www.census.gov/, 2013.

\bibitem{Andersen02}
D.~G. Andersen, N.~Feamster, S.~Bauer, and H.~Balakrishnan.
\newblock Topology inference from {BGP} routing dynamics.
\newblock In {\em Internet Measurement Workshop}, pages 243--248, 2002.

\bibitem{wdi2012}
W.~Bank.
\newblock {\em World Development Indicators 2012}.
\newblock The World Bank, 2012.

\bibitem{akamai12}
D.~Belson.
\newblock {The State of the Internet, 2nd Quarter, 2012 Report}.
\newblock {\em Akamai}, 5(2), 2012.

\bibitem{CFKW11}
N.~Czernich, O.~Falck, T.~Kretschmer, and L.~Woessmann.
\newblock Broadband infrastructure and economic growth*.
\newblock {\em The Economic Journal}, 121(552):505--532, 2011.

\bibitem{dbip}
{DB-IP}.
\newblock {IP-City Database}.
\newblock http://db-ip.com/, 2012.

\bibitem{DD11}
A.~Dhamdhere and C.~Dovrolis.
\newblock {Twelve Years in the Evolution of the Internet Ecosystem}.
\newblock {\em IEEE/ACM Transactions on Networking}, 19(5):1420--1433, sep
  2011.

\bibitem{DS11}
A.~Dhamdhere and C.~Dovrolis.
\newblock Twelve years in the evolution of the internet ecosystem.
\newblock {\em IEEE/ACM Trans. Netw.}, 19(5), Oct. 2011.

\bibitem{netacuity}
{Digital Envoy}.
\newblock {NetAcuity Edge}.
\newblock $http://www.digital-element.com/our\_technology/edge.html$, 2010.

\bibitem{WebDimes}
DIMES.
\newblock {Distributed Internet Measurements and Simulations}.
\newblock {http://www.netdimes.org/}.

\bibitem{FSZ-Comnet12}
D.~Feldman, Y.~Shavitt, and N.~Zilberman.
\newblock A structural approach for {PoP} geolocation.
\newblock {\em Computer Networks}, 56(3):1029--1040, 2012.

\bibitem{geobytes}
{Geobytes}.
\newblock {GeoNetMap}.
\newblock $http://www.geobytes.com/$, 2010.

\bibitem{HJCS10}
M.~A. Hameed, A.~Jabbar, E.~K. \c{C}etinkaya, and J.~P. Sterbenz.
\newblock {Deriving Network Topologies from Real World Constraints}.
\newblock In {\em IEEE GLOBECOM Workshop on Complex and Communication Networks
  (CCNet)}, pages 415--419, December 2010.

\bibitem{ip2location}
{Hexsoft Development}.
\newblock {IP2Location}.
\newblock http://www.ip2location.com, 2010.

\bibitem{hostip}
{hostip.info}.
\newblock {hostip.info}.
\newblock http://www.hostip.info, 2010.

\bibitem{IPligence}
{IPligence}.
\newblock {IPligence Max}.
\newblock http://www.ipligence.com, 2010.

\bibitem{K12}
J.~Kolko.
\newblock Broadband and local growth.
\newblock {\em Journal of Urban Economics}, 71(1):100--113, 2012.

\bibitem{K09}
P.~Koutroumpis.
\newblock The economic impact of broadband on growth: A simultaneous approach.
\newblock {\em Telecommunications Policy}, 33(9):471--485, 2009.

\bibitem{LBCM03}
A.~Lakhina, J.~W. Byers, M.~Crovella, and I.~Matta.
\newblock {On the geographic location of Internet resources}.
\newblock {\em Selected Areas in Communications, IEEE Journal on},
  21(6):934--948, Aug. 2003.

\bibitem{LMHSCV11}
S.~Laki, P.~M\'{a}tray, P.~H\'{a}ga, T.~Seb\"{o}k, I.~Csabai, and G.~Vattay.
\newblock Spotter: A model based active geolocation service.
\newblock In {\em Proceedings of IEEE INFOCOM}, Shanghai, China, 2011.

\bibitem{madhyastha08thesis}
H.~V. Madhyastha.
\newblock An information plane for internet applications.
\newblock Thesis, University of Washington, 2008.

\bibitem{Madhyastha06}
H.~V. Madhyastha, T.~Anderson, A.~Krishnamurthy, N.~Spring, and
  A.~Venkataramani.
\newblock A structural approach to latency prediction.
\newblock In {\em Proceedings of the 6th ACM SIGCOMM conference on Internet
  measurement (IMC'06)}, pages 99--104, 2006.

\bibitem{MHLVC12}
P.~M\'{a}tray, P.~H\'{a}ga, S.~Laki, G.~Vattay, and I.~Csabai.
\newblock On the spatial properties of internet routes.
\newblock {\em Comput. Netw.}, 56(9):2237--2248, June 2012.

\bibitem{maxmind}
{MaxMind LLC}.
\newblock {GeoIP}.
\newblock http://www.maxmind.com, 2010.

\bibitem{maxmind_worldcities}
{MaxMind LLC}.
\newblock {Free World Cities Database}.
\newblock \\$http://www.maxmind.com/en/worldcities$, 2012.

\bibitem{PV04}
R.~Pastor-Satorras and A.~Vespignani.
\newblock {\em Evolution and Structure of the Internet: A Statistical Physics
  Approach}.
\newblock Cambridge University Press, New York, NY, USA, 2004.

\bibitem{tv-penetration}
{Point Topic LTD.}
\newblock {IPTV Statistics - Market Analysis}.
\newblock
  \\$http://point-topic.com/wp-content/uploads/2013/02/$\\$Sample-Report-Global-IPTV-Statistics-Q2-2012.pdf$,
  2012.

\bibitem{RPK01}
S.~Rinaldi, J.~Peerenboom, and T.~Kelly.
\newblock Identifying, understanding, and analyzing critical infrastructure
  interdependencies.
\newblock {\em Control Systems, IEEE}, 21(6):11--25, 2001.

\bibitem{RW01}
L.-H. Roller and L.~Waverman.
\newblock Telecommunications infrastructure and economic development: A
  simultaneous approach.
\newblock {\em American Economic Review}, 91(4):909--923, 2001.

\bibitem{ipums}
S.~Ruggles, J.~T. Alexander, K.~Genadek, R.~Goeken, M.~B. Schroeder, and
  M.~Sobek.
\newblock {\em {Integrated Public Use Microdata Series: Version 5.0
  [Machine-readable database]}}.
\newblock Minneapolis: University of Minnesota, 2010.

\bibitem{SFKKC09}
S.~Shakkottai, M.~Fomenkov, R.~Koga, D.~Krioukov, and K.~Claffy.
\newblock Evolution of the internet as-level ecosystem.
\newblock In {\em Complex Sciences}, volume~5 of {\em Lecture Notes of the
  Institute for Computer Sciences, Social Informatics and Telecommunications
  Engineering}, pages 1605--1616. 2009.

\bibitem{SZ_JSAC11}
Y.~Shavitt and N.~Zilberman.
\newblock A geolocation databases study.
\newblock {\em IEEE Journal on Selected Areas in Communications}, 29(9), 2011.

\bibitem{SZ12_arabian}
Y.~Shavitt and N.~Zilberman.
\newblock Arabian nights: measuring the arab internet during the 2011 events.
\newblock {\em Network, IEEE}, 2012.

\bibitem{shavitt2013improving}
Y.~Shavitt and N.~Zilberman.
\newblock Improving {IP} geolocation by crawling the internet {PoP} level
  graph.
\newblock In {\em IFIP Networking Conference, 2013}, pages 1--9. IEEE, 2013.

\bibitem{Spring02journal}
N.~T. Spring, R.~Mahajan, D.~Wetherall, and T.~E. Anderson.
\newblock Measuring {ISP} topologies with {Rocketfuel}.
\newblock {\em IEEE/ACM Transactions on Networking}, 12(1):2--16, 2004.

\bibitem{broadband-penetration}
{TeleGeography}.
\newblock {IPTV Broadband Penetration Reaches 15 Percent, Growth Prospects are
  Patchy}.
\newblock \\$http://www.telegeography.com/press/marketing-emails/2012/06/20/
  iptv-broadband-penetration-reaches-15-percent-growth-prospects-are-patchy/index.html$,
  2012.

\bibitem{us-age}
{United States Census Bureau}.
\newblock {Metropolitan Statistical Areas--Population by Age: 2010}.
\newblock \\$http://www.census.gov/compendia/statab/2012/tables/\\12s0022.xls$,
  2012.

\bibitem{us-race}
{United States Census Bureau}.
\newblock {Metropolitan Statistical Areas--Population by Race: 2010}.
\newblock \\$http://www.census.gov/compendia/statab/2012/tables/\\12s0023.xls$,
  2012.

\bibitem{us-bea}
{United States Department of Commerce}.
\newblock {Bureau of Economic Analysis}.
\newblock http://www.bea.gov/, 2013.

\bibitem{us-transportation}
{United States Department of Transportation}.
\newblock {Bureau of Transportation Statistics}.
\newblock http://www.rita.dot.gov/bts/, 2013.

\bibitem{bea-income}
{US Department of Commerce, Bureau of Economic Analysis}.
\newblock {Personal Income and Per Capita Personal Income by Metropolitan Area,
  2009 - 2011)}.
\newblock
  $http://www.bea.gov/newsreleases/regional/lapi/\\lapi\_newsrelease.htm$,
  2012.

\bibitem{bea-gdp}
{US Department of Commerce, Bureau of Economic Analysis}.
\newblock {Real GDP by Metropolitan Area (millions of chained 2005 dollars)}.
\newblock
  $\\http://www.bea.gov/iTable/iTable.cfm?reqid=70\&step=1\&isuri=1\&acrdn=2\#$,
  2013.

\bibitem{bts-congestion}
{US Department of Transportation, Bureau of Transportation Statistics}.
\newblock {Highway Congestion in the 50 Largest Urban Areas: 2010}.
\newblock
  \\$http://apps.bts.gov/publications/state\_transportation\\\_statistics/state\_transportation\_statistics\_2011/html/\\table\_05\_05.html$,
  2011.

\bibitem{bts-airports}
{US Department of Transportation, Bureau of Transportation Statistics}.
\newblock {Passengers Boarded at the Top 50 U.S. Airports}.
\newblock
  \\$http://www.rita.dot.gov/bts/sites/rita.dot.gov.bts/files/\\publications/national\_transportation\_statistics/html/\\table\_01\_44.html$,
  2013.

\bibitem{bts-gateways}
{US Department of Transportation, Bureau of Transportation Statistics}.
\newblock {Top U.S. Foreign Trade Freight Gateways by Value of Shipments}.
\newblock
  \\$http://www.rita.dot.gov/bts/sites/rita.dot.gov.bts/files/\\publications/national\_transportation\_statistics/html/\\table\_01\_51.html$,
  2013.

\bibitem{VFV10}
S.~Vinciguerra, K.~Frenken, and M.~Valente.
\newblock The geography of internet infrastructure: An evolutionary simulation
  approach based on preferential attachment.
\newblock {\em Urban Studies}, 47(9):1969--1984, 2010.

\bibitem{WL10}
X.~Wang and D.~Loguinov.
\newblock Understanding and modeling the internet topology: economics and
  evolution perspective.
\newblock {\em IEEE/ACM Trans. Netw.}, 18(1):257--270, Feb. 2010.

\bibitem{YJB02}
S.-H. Yook, H.~Jeong, and A.-L. Barab\'{a}si.
\newblock {Modeling the Internet's large-scale topology}.
\newblock {\em Proceedings of the National Academy of Sciences},
  99(21):13382--13386, Oct. 2002.

\bibitem{yoshida09}
K.~Yoshida, Y.~Kikuchi, M.~Yamamoto, Y.~Fujii, K.~Nagami, I.~Nakagawa, and
  H.~Esaki.
\newblock Inferring {PoP}-level {ISP} topology through end-to-end delay
  measurement.
\newblock In {\em Proceedings of the 10th international conference on Passive
  and Active Measurement (PAM'09)}, volume 5448, pages 35--44, 2009.

\bibitem{zilberman13thesis}
N.~Zilberman.
\newblock {\em {The Internet PoP Level Graph}}.
\newblock PhD thesis, Tel-Aviv University, 2013.

\end{thebibliography}
\end{small}
\label{last-page}

\end{document}